\documentstyle[11pt,psfig]{article}

\makeatletter

\newcommand{\etal}{{\it et al.} }

\makeatother

\begin{document}

\vspace{1.0cm} \textbf{\large THE BROAD BAND SPECTRAL PROPERTIES OF BINARY
X--RAY PULSARS}
\footnote[4]{Accepted for publication in {\it Advances in Space
Research} - Proceedings of 32nd COSPAR Scientific Assembly - Symposium
E1.1: ``Broad Band X-ray Spectroscopy of Cosmic Sources''}{\large \par}

\vspace{1.0cm}

D. Dal Fiume\( ^{1} \), M. Orlandini\( ^{1} \), S. Del Sordo\( ^{2} \), F.
Frontera\( ^{1,3} \), T. Oosterbroek\( ^{4} \), E. Palazzi\( ^{1} \),
A. N. Parmar\( ^{4} \), S. Piraino\( ^{2} \), A. Santangelo\( ^{2} \) and
A. Segreto\( ^{2} \)

\vspace{1.0cm} \( ^{1} \)\textit{TeSRE, Consiglio Nazionale delle Ricerche,
via Gobetti 101, I--40129 Bologna, Italy}\\
\( ^{2} \)\textit{IFCAI, Consiglio Nazionale delle Ricerche,
via La Malfa 153, I-90146 Palermo, Italy}\\
\( ^{3} \)\textit{Dipartimento di Fisica, Universit\'{a} di Ferrara, via
Paradiso 1, I--44100 Ferrara, Italy}\\
\( ^{4} \)\textit{ESA--SSD, ESTEC, Keplerlaan 1, 2200 AG Noordwijk,
The Netherlands}\\

\vspace{0.5cm}

\section*{ABSTRACT}

The X--ray telescopes on board BeppoSAX are an optimal set of instruments to
observe bright galactic binary pulsars. These sources emit very hard and quite
complex X--ray spectra that can be accurately measured with BeppoSAX between
0.1 and 200 keV. A prototype of this complexity, the source Her X--1, shows
at least seven different components in its spectrum. A broad band measure is
therefore of paramount importance to have a thorough insight into the physics
of the emitting region. Moreover the detection of cyclotron features,
when present, allows a direct and highly significant measure of the
magnetic field intensity in the emission region.

In this paper we briefly report the results obtained with BeppoSAX on this
class of sources, with emphasis on the detection and on the measured
properties of the cyclotron lines.

\section{INTRODUCTION}

X--ray pulsars are studied since more than 25 years but still we are far
from understanding the details of the mechanisms that produce the emission
of their complex and variable spectra. The overall scenario was depicted with
bright accuracy soon after their discovery (Davidson and Ostriker 1973). As
the observational data became of better and better quality, it clearly emerged
that the X-ray spectra emitted from these sources are quite complex. The spectra
of almost all sources are characterized by a very hard power--law like emission
in the 1--10 keV band (photon index \( \alpha  \) between 0 and -1) and by
a high energy cutoff, approximately exponential, starting between 10 and 20
keV.

While the task of understanding and convincingly modeling the X--ray emission
is exceedingly difficult, the final reward to a success in this field is rather
appealing. The emitting neutron stars are by themselves relativistic objects.
Moreover the magnetic field, needed to channel the accreting matter onto the
surface hot spots from which we observe the pulsed emitted radiation,
is relativistic.
Therefore these systems are an ideal laboratory to test relativistic effects
by observational means.

In this paper
we report the observations of a small sample of bright X--ray pulsars,
describing the observational results both on the broad band spectrum and
on the cyclotron line features.

\section{OBSERVATIONS AND RESULTS}

During its first two years of operational life BeppoSAX observed all the
bright persistent X--ray pulsars and one recurrent transient pulsar (GS
1843+00).
The BeppoSAX observations of bright X--ray pulsars are listed in Table 1.

%
{\centering 
{\bf Table 1}\\
\medskip
\begin{tabular}{|lc|lc|}
\hline 
Source name&  \multicolumn{1}{l|}{Total elapsed time (ks)}&
Source name&  \multicolumn{1}{l|}{Total elapsed time (ks)}\\
\hline 
SMC X-1& 75 &4U 1538-52$^{(*)}$&  234\\
X Persei& 227&4U 1626-67$^{(*)}$& 172\\
LMC X-4& 163&GX 1+4& 148\\
Vela X-1$^{(*)}$& 625&Her X-1$^{(*)}$&  1036\\
Cen X-3$^{(*)}$& 97&GX 301-2& 450 \\
GS 1843+00& 46&4U 1907+09$^{(*)}$& 230\\
\hline
\multicolumn{4}{l}{ }\\
\multicolumn{4}{l}{(*): cyclotron features in the spectra}\\
\end{tabular}\par}
%

A straightforward observational goal is to obtain a possibly simple and
unique model
spectrum to fit data from different sources. Restricting ourselves
to a purely heuristic, and thence phenomenological, approach, and using
the more common spectral models (e.g. White et al. 1983, Mihara
1995), we are not able to obtain satisfactory fits covering the entire
BeppoSAX energy band for all the observed sources using only one
modelization.

In some cases, like Her X--1 and possibly 4U1626--67, additional
components at energies $\leq$ 1 keV must be added. In other cases, like
Vela X--1 and GX 301--2, a highly variable source intensity is coupled
with huge variations in intrinsic opacity at low energies ($\leq$ 2 keV)
due to a highly varying amount of intervening matter along the line of
sight. These additional spectral components are variable with time. In
the case of GX 301--2 the observational results are further complicated
by the substantial value of this variable absorption at low energies,
with an equivalent hydrogen column N$_H \sim 10^{23} - 10^{24}$.

This difficulty may seem purely phenomenological, as intrinsic unabsorbed
X--ray spectra emerging from the accreting zones onto the neutron star
surface may quite reasonably differ due to differences in mass accretion
rate, NS mass, NS magnetic field, eventually NS magnetic field
geometry. However this ambiguity must be solved in order to have
reliable direct measures of relevant physical quantities. In fact the
other spectral components that seem quite frequent in the spectra of
X--ray pulsars as measured by BeppoSAX are cyclotron features.
These features appear substantially broad and cluster between 20 and 60 keV.
Adding other previous measurements to the BeppoSAX ones, one can conclude
that up to now the cyclotron line energies range between 10 and
100 keV, as the lower cyclotron resonance energy was measured at 11--12
keV in 4U 0115+63 (White et al. 1983; Nagase et al. 1991) and the higher
was measured at 100--115 keV in A0535+26 (Grove et al. 1995). \it As
these features are broad, they are strongly coupled to the continuum
shape and intensity and therefore the parameters that characterize each
feature in each given source are slightly but clearly dependent on the
choice of the continuum spectral model.\rm\\
This point must be kept in mind while looking at Table 2, where we
report the results of broad band fits to the X--ray pulsars spectra
using only BeppoSAX published data.

Apart from these cautionary remarks, 
as the differences between source spectra in the sample are a
obvious difficulty in the use of a single model, some
commonalities do emerge. In all the measured spectra we can easily
identify some characteristic components, already suggested more than
fifteen years ago by White et al (1983): a power law at energies below
10 keV; a cutoff between $\sim 6-7$ and $\sim 20-30$ keV; a high energy
tail up to 100 keV (and more for the harder sources).
While the details in the modeling of the single sources differ, these
components are stably present in all the sources we are discussing in
this report.

In Table 2 we summarize the results from BeppoSAX on these common
components. We caution that this is a first order approach to source
spectra modelization, as it is often necessary to further complicate
these components in order to get reasonable values of $\chi^2$ from the
spectral fits. \\
The count rate broad band spectra of these sources are shown in Figure
1 (references as in Table 2).

Up to now,
cyclotron lines are present in 6 out of the 12 bright X--ray pulsars
observed by SAX, including 4U1538--52 that is not in this sample.
The observations of GX 301--2 are being analyzed.\\
It is remarkable the fact that the two hardest sources
in this small sample (GX 1+4 and GS 1843+00) do not show any feature at
least up to 100-200 keV. The other sources in whose spectra
cyclotron features were detected are the recurrent transient pulsars
A0535+26 (Grove et al. 1995), 4U 0115+63 (White et al. 1983; Nagase et
al. 1991), X0331+53 (Makishima et al. 1990), Cep X--4 (Mihara et al.
1991). These sources were quiescent during the first two years of
BeppoSAX operational life.\\
\begin{center}
{\bf Table 2}\\
\smallskip
\begin{tabular}{|lllllll|}
\hline
Source name & Power law & High en.& High en.& Cycl. line&
Cycl. line& Model$^1$\\
            & spectral index& cutoff & folding 
& centroid & FWHM & \\
   &   &   (keV)  &  (keV)  &  (keV) & (keV) & \\
\hline
 & & & & & & \\
 Cen X--3 &  1.21$\pm$0.01 & 14.9 $\pm$ 0.2 & 12.1$\pm$0.9 & 27.9$\pm$
 0.5 & 9.8$\pm$ 1.9& (1) \\
 4U 1626--67 & 0.86$\pm$0.01 & 19.6$\pm$0.5 & 10.6$\pm$0.7 & 37$\pm$1 &
 7.05$\pm$2.35 & (2) \\
 4U 1907+09$^2$ & 1.27$\pm$0.01 & 12.0$\pm$0.3 & 12.0$\pm$0.3 & 39.4$\pm$0.6
 & 8.5$\pm$1.6& (2)  \\
 Her X--1 & 0.884$\pm$.003 & 24.2$\pm$0.2 & 14.8$\pm$0.4 & 42.1$\pm$0.3
 & 14.9$^{+1.25}_{-1.0}$& (1)  \\
 Vela X--1$^2$ & (NPEX) & (NPEX) & (NPEX) & 
 54.4$^{+1.5}_{-0.2}$&17$^{+3}_{-2}$ & (3)\\
 GX1+4$^3$ & 1.0 & 25$\pm$ 2& 29$\pm$ 1& no line & no line & (2)\\
 GS 1843+00$^3$ & 0.34$\pm$ 0.04 & 5.95$\pm$ 0.45 & 18.4$\pm$ 0.6&
 no line & no line & (2)\\
 \hline
 \multicolumn{6}{l}{ }\\
 \multicolumn{6}{l}{{ }$^1$ \footnotesize The models for the continuum
 used in the reference papers are: \bf(1)\ \rm broken power law plus high}\\
 \multicolumn{6}{l}{{ }~~~\footnotesize 
 energy cutoff; \bf(2)\ \rm power law plus high energy cutoff; \bf(3)\ \rm
 NPEX (Mihara 1995)}\\
 \multicolumn{6}{l}{{ }$^2$ \footnotesize In these sources it is
 suspected the
 presence of cyclotron features both at the fundamental energy}\\
 \multicolumn{6}{l}{{ }~~~\footnotesize
 and at the first harmonic}\\
 \multicolumn{6}{l}{{ }$^3$ \footnotesize No cyclotron line in the
 X--ray spectrum}\\
 \multicolumn{6}{l}{{ }$^4$ \footnotesize References are: Cen X--3,
 Santangelo et al. 1998; 4U 1626--67, Orlandini et al. 1998a; 4U1907+09,}\\
 \multicolumn{6}{l}{{~~~}\footnotesize
 Cusumano et al. 1998; Her X--1, Dal Fiume et al. 1998; Vela X--1,
 Orlandini et al. 1998b; GS 1843+00,}\\
 \multicolumn{6}{l}{{~~~}\footnotesize
 Piraino et al. 1998, 1999; GX1+4, Israel et al. 1998}\\
 \end{tabular}
\end{center}
A major observational topic in these last years has become the measure
of multiplicity in the cyclotron line features. In Table 2 we marked
with an asterisk the two sources for which there is some evidence of the
presence of two features that may tentatively be identified as the
fundamental resonance and the first harmonic.
In both cases the fundamental should be half the energy reported in
Table 2.\\
From a purely
observational point of view, still some ambiguity remains. In both cases
the fundamental should be of quite low equivalent width, and it is
located near or exactly at the high energy cutoff. This is a quite
relevant complication from a purely technical point of view. In fact,
while the deeper feature located at higher energy ($\sim$55 keV for Vela
X--1 and $\sim$39 keV for 4U1907+09) is superposed to a continuum with a
well defined slope, that "after" the feature clearly recovers the shape
shown "before", in the case of the possible feature at $\sim$27 keV for
Vela X--1 and at $\sim$19 keV for 4U1907+09 the shape of the continuum
is rapidly evolving with energy, and it is also reasonable that the
exponential--like shape of the cutoff is only a purely phenomenological
approach. This empirical model has also the disadvantage to have a
jump in its first derivative at the cutoff energy.\\
In this sense a sensible attempt to model the shape of the broad band
continuum of X--ray pulsars without the presence of a second order
discontinuity of
the power--law--plus--cutoff model of White et al. (1983) was attempted
by Mihara (1995). His results convincingly show that some more
smooth functionals, like the NPEX model (negative plus positive
power law plus exponential), give quite satisfactory fits to the X--ray
pulsars in the GINGA bandpass. However the extension of the measures to
a broader energy range with BeppoSAX shows that unfortunately this
functional must be modified in order to get acceptable values of
$\chi^2$ from the fits.\\
\centerline{\psfig{file=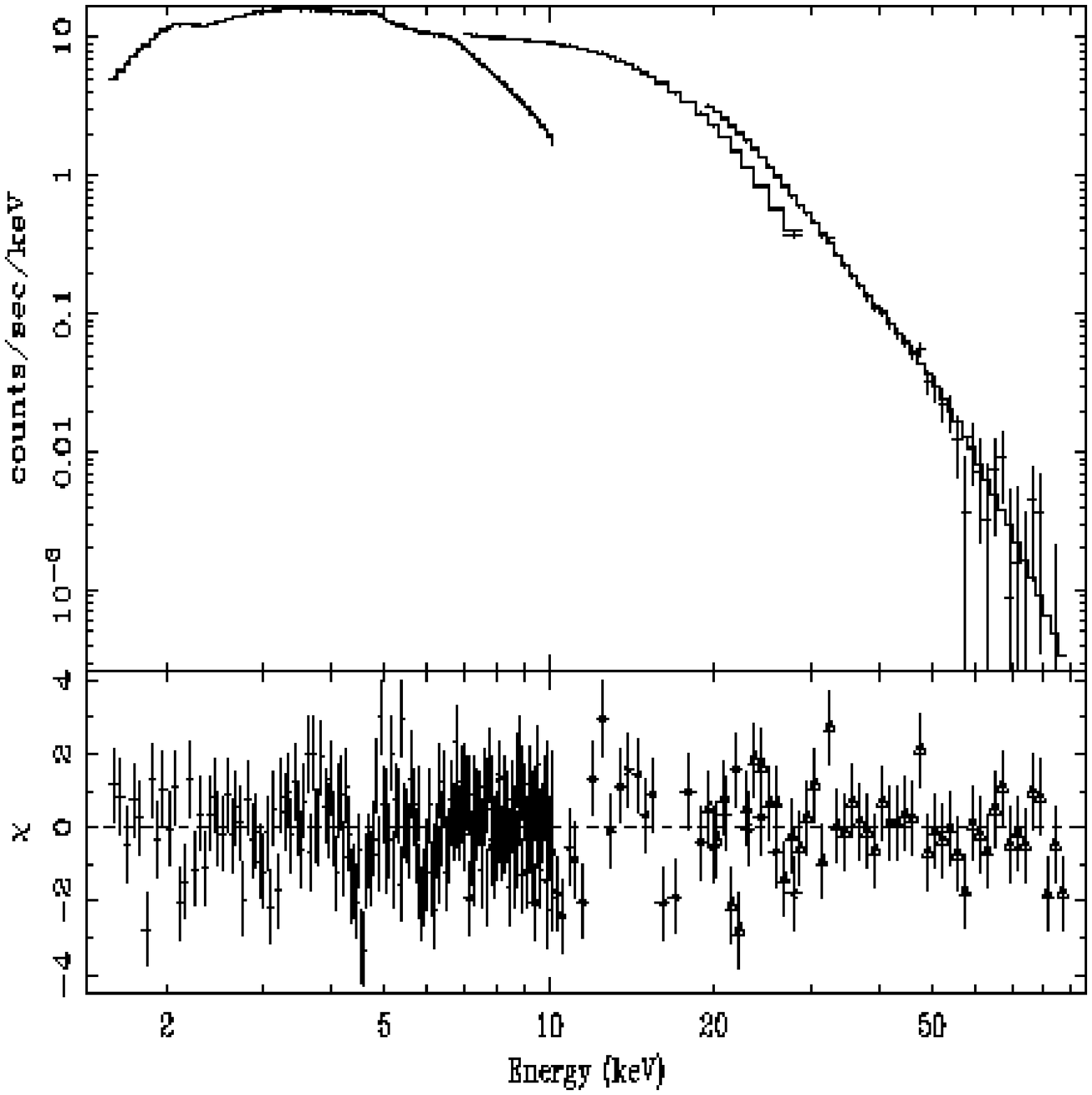,width=0.45\textwidth,height=10.5truecm}
\hfill
\psfig{file=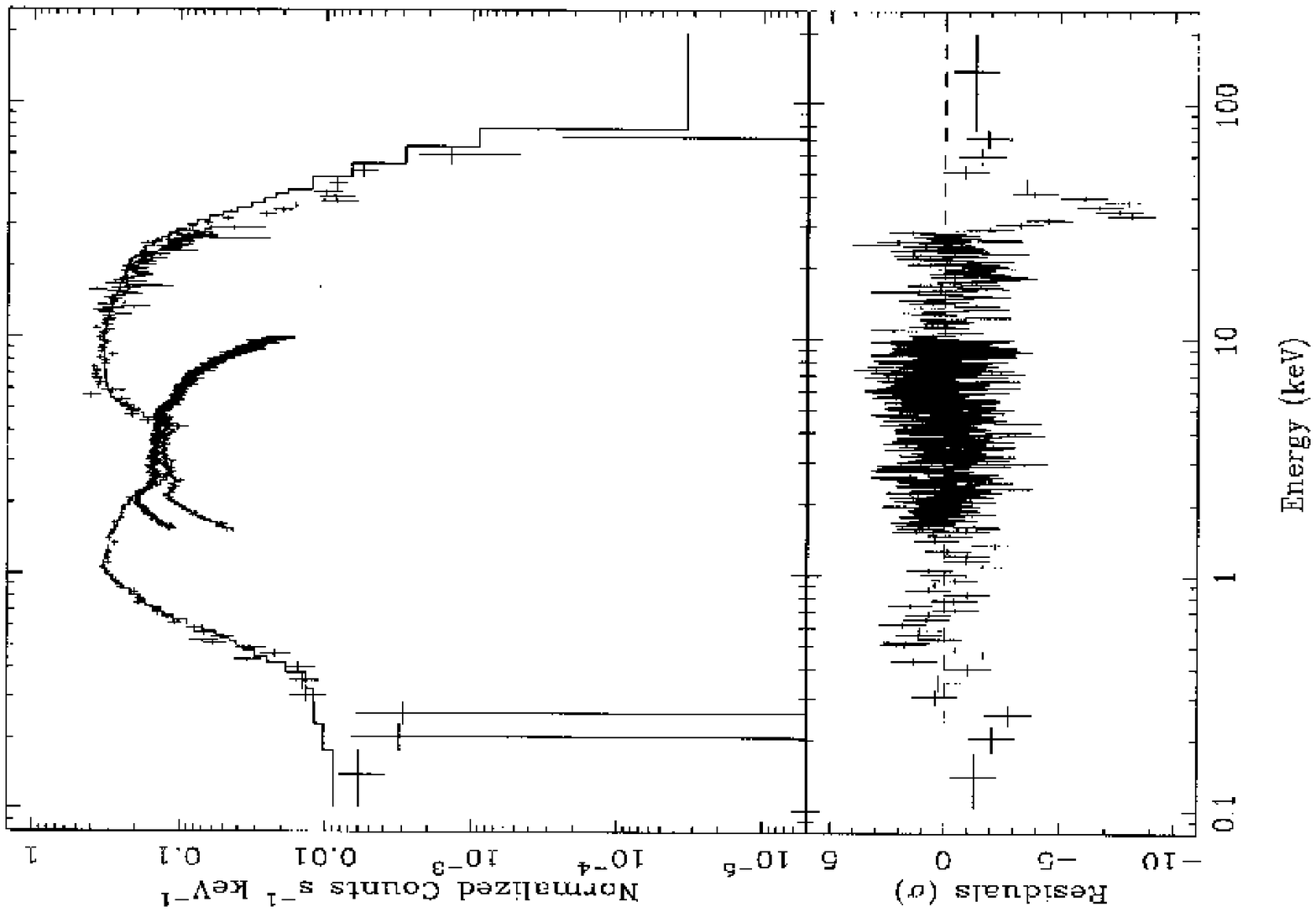,width=0.45\textwidth,height=10.5truecm,angle=270}}

\centerline{\psfig{file=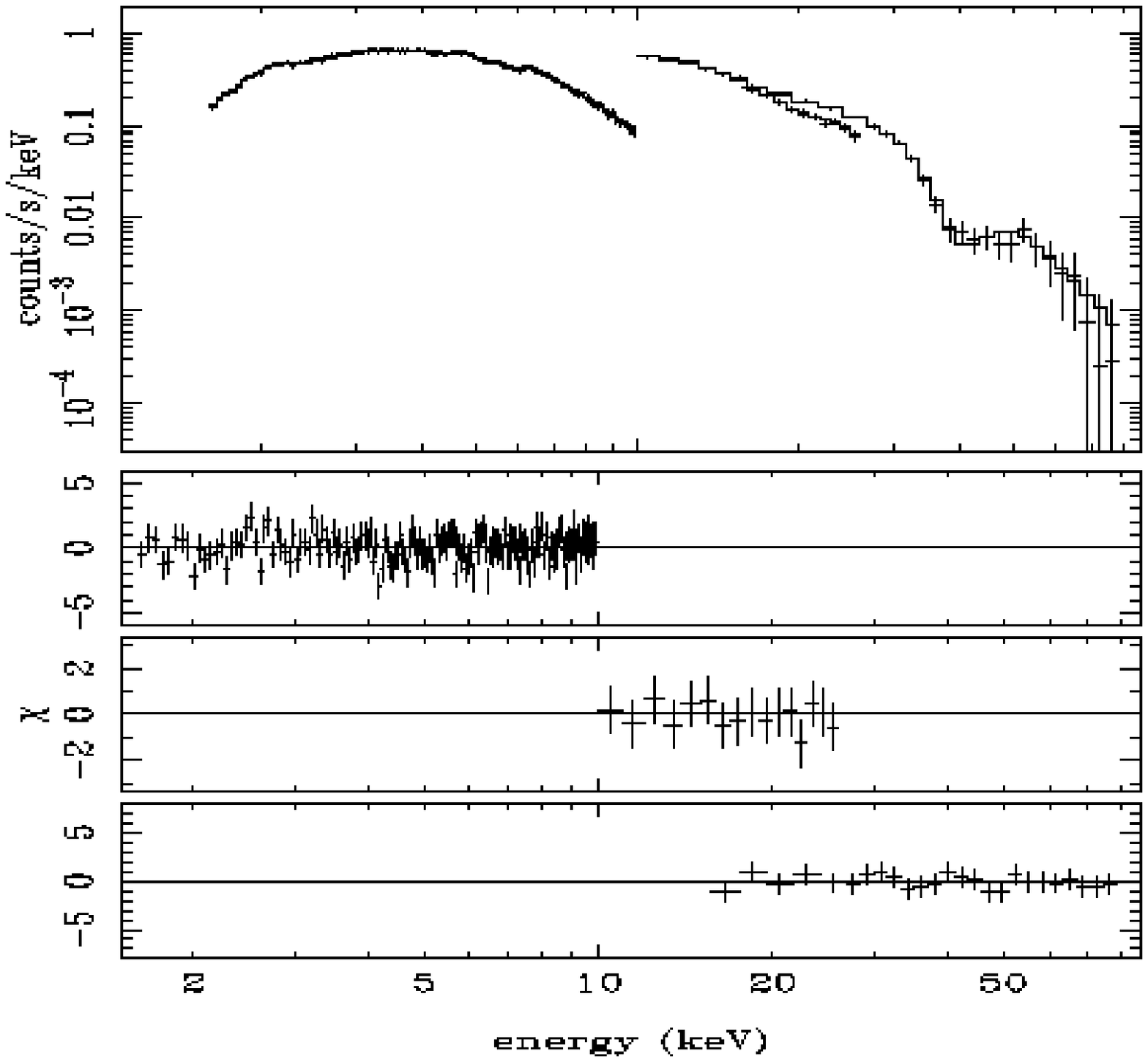,width=0.45\textwidth,height=10.5truecm,angle=270}
\hfill
\psfig{file=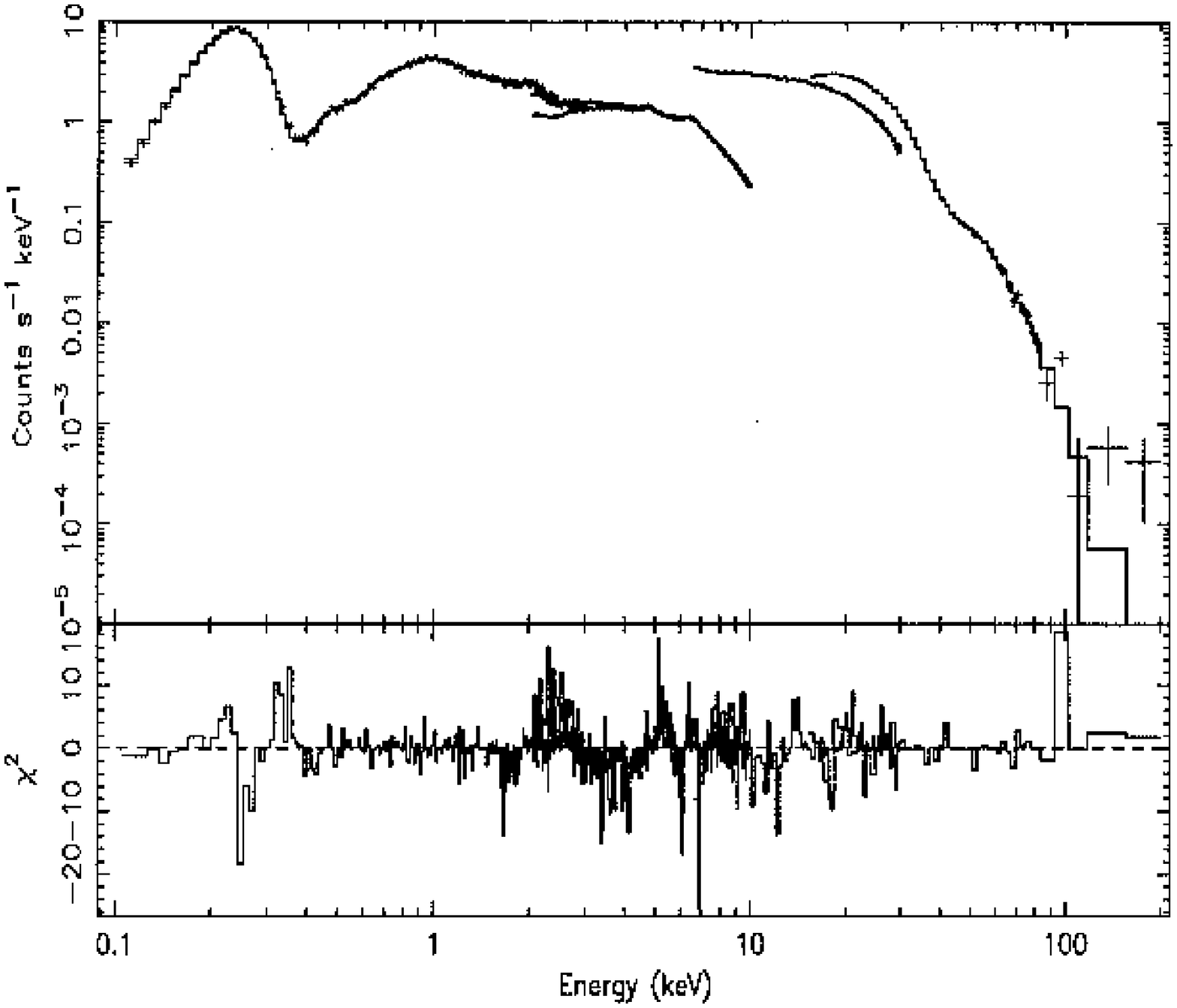,width=0.45\textwidth,height=10.5truecm,angle=270}}
\begin{center}
\begin{minipage}[t]{18cm}
\bf Figure 1a:\rm BeppoSAX count rate spectra of Cen X--3, 4U 1626--67,
Her X--1 and 4U1907+09 (clockwise from top left)
\end{minipage}
\end{center}

\centerline{\psfig{file=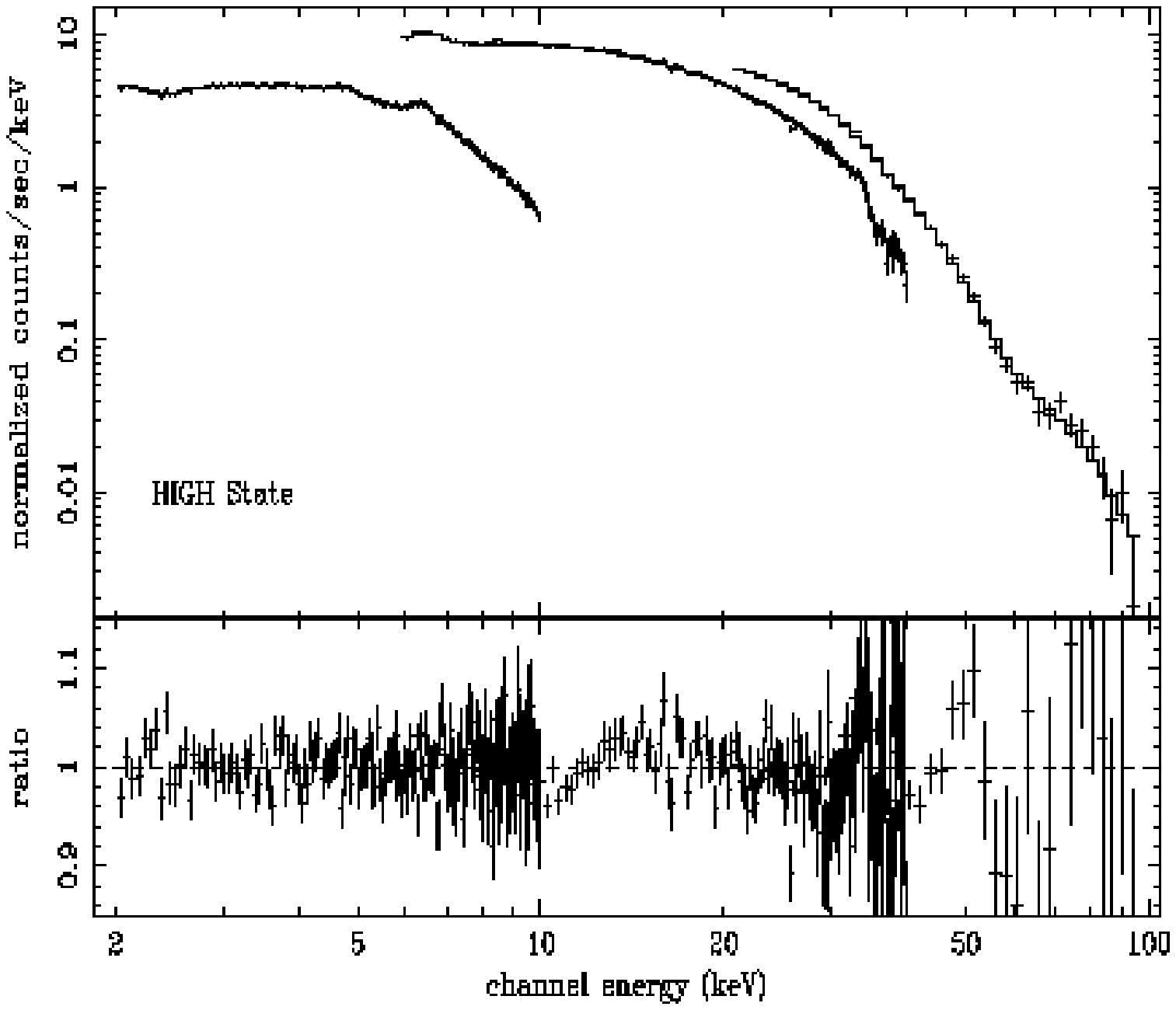,width=0.45\textwidth,height=10.95truecm,angle=270}
\hfill
\psfig{file=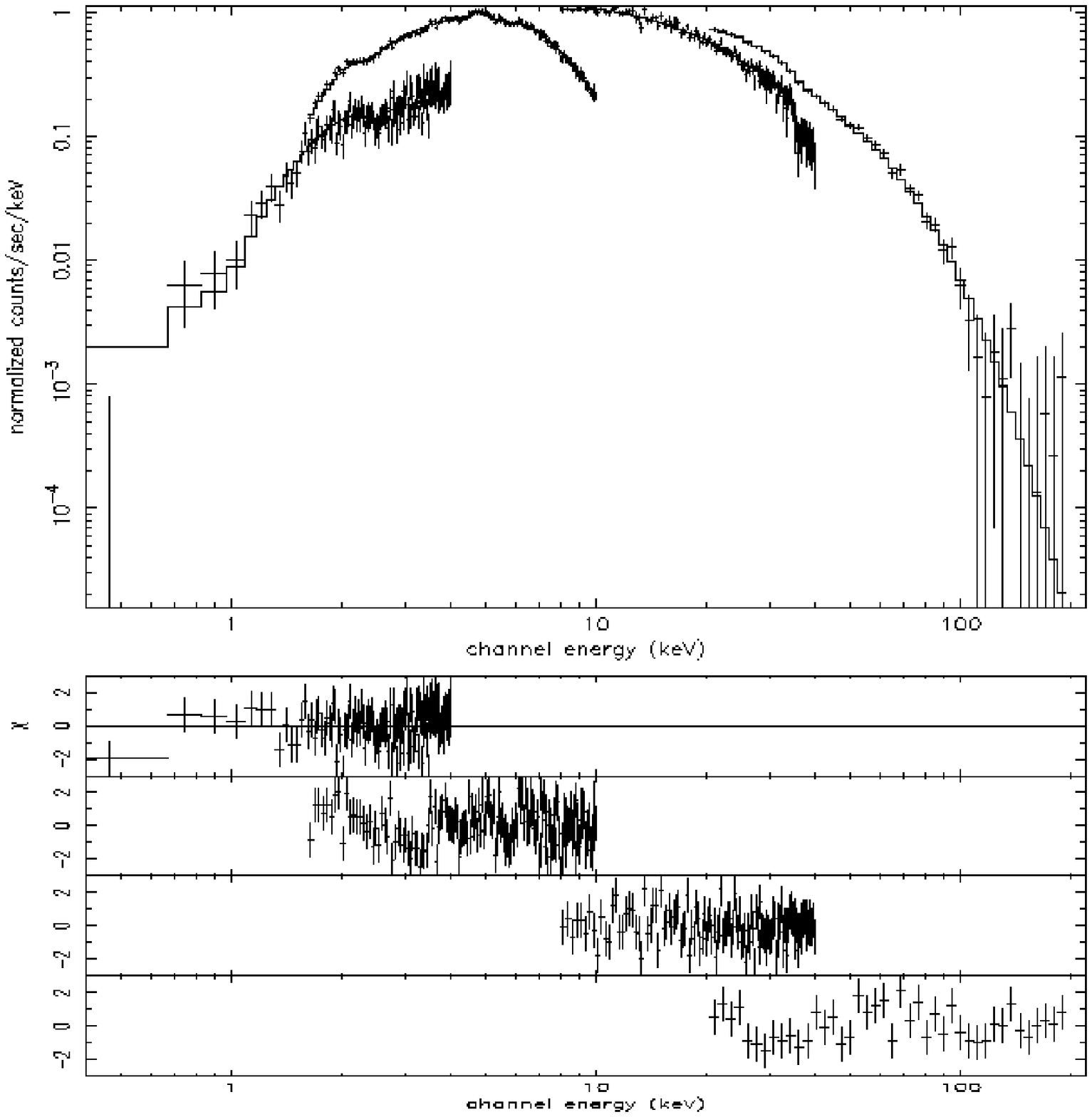,width=0.45\textwidth,height=10.95truecm,angle=270}
}

\centerline{
\psfig{file=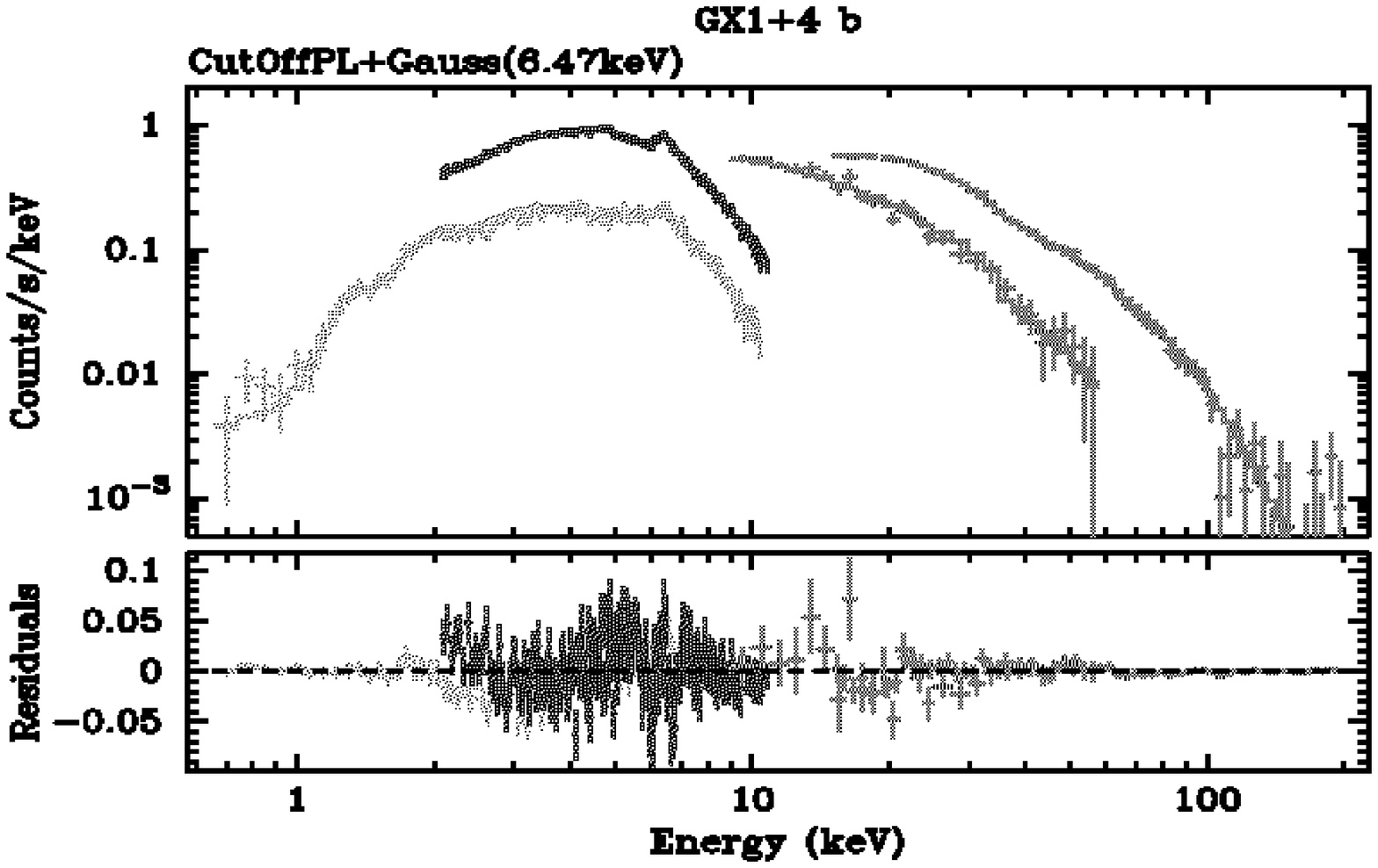,width=0.45\textwidth,height=10.95truecm}
}
\begin{center}
\begin{minipage}[t]{18cm}
\bf Figure 1b:\rm BeppoSAX count rate spectra of Vela X--1 (top left), GS
1843+00 (top right) and GX1+4 (bottom)
\end{minipage}
\end{center}
\clearpage

A reliable method to enhance the presence of features in
the X--ray spectra of cosmic sources is the so called "Crab ratio".
This method was extensively used by Mihara (1995) in his survey of
X--ray pulsars observed with GINGA. This ratio is simply the ratio
between the source count rate spectrum and the count rate spectrum of
the Crab Nebula. As this second spectrum is known, with great accuracy,
to be free of features and to be modeled at first order with a power law
in a very broad energy range, this ratio is quite well suited to enhance
the presence of features in the spectrum (iron line, iron K edge,
cyclotron features). Furthermore the ratio is in first approximation
independent from the calibration of the instrument.\\
We added to this
method a further step, that gives easily readable and comparable plots.
After this ratio on count rate, we multiply by a E$^{-2.1}$ power law,
that is the functional form of the Crab Nebula spectrum, and we
divide by the functional describing the continuum shape of the source.
The procedure is described in Figure 2, where we plot the result of
each different step used to obtain the final result in the case of
4U1626--67.\\
\centerline{\psfig{file=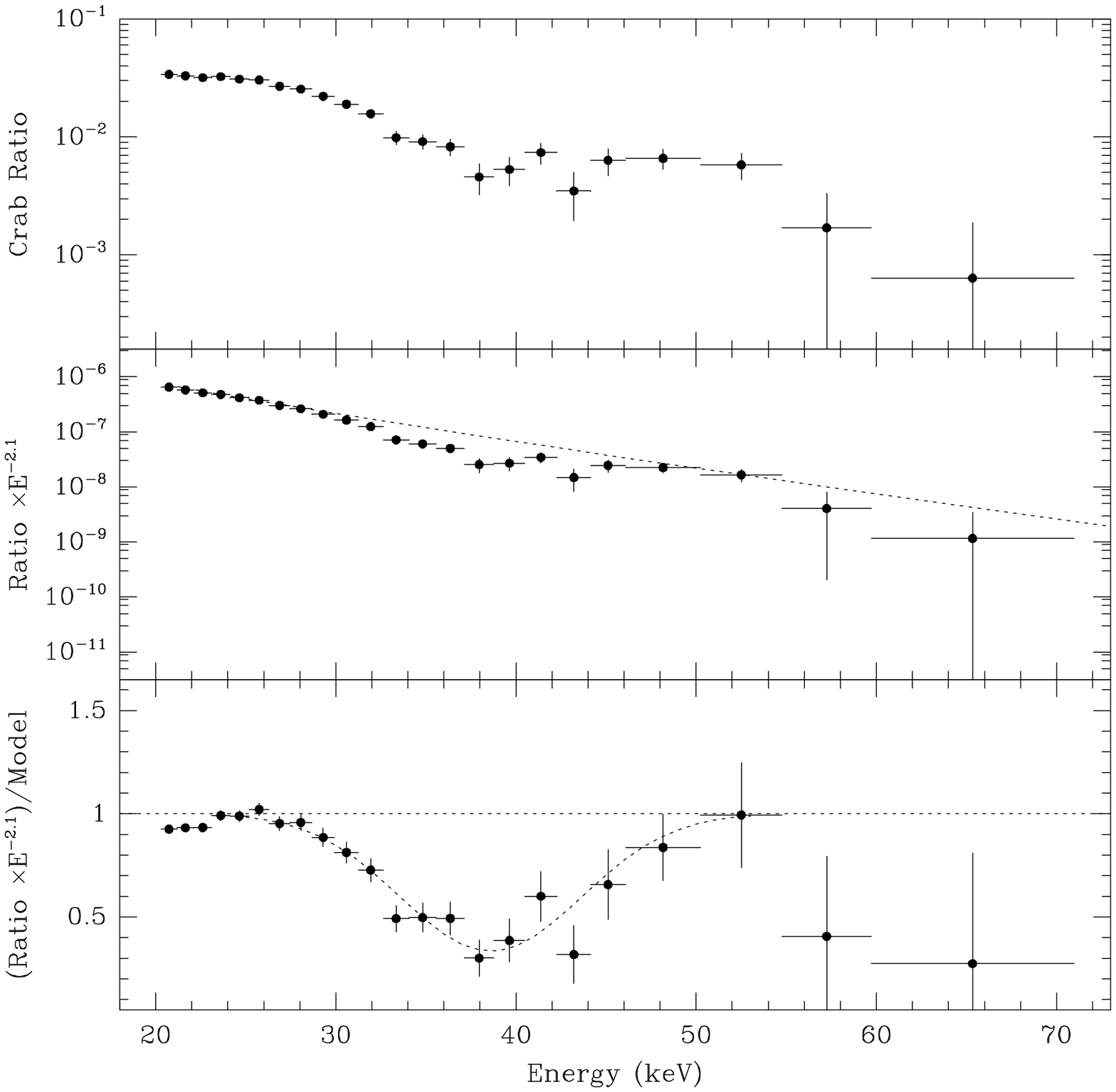,height=10truecm}}
\begin{center}
\begin{minipage}[t]{0.8\textwidth}
{\bf Figure 2:\rm first panel: ratio between the count rate spectrum of
the Crab Nebula and the count rate spectrum of 4U1626--67; second panel:
the ratio multiplied by a function describing the Crab photon spectrum
($E^{-2.1}$); third panel: the normalized ratio obtained dividing the
result plotted in the second panel by a function describing the continuum
of 4U1626--67}
\end{minipage}
\end{center}
In Figure 3 we plot the normalized ratios for the X--ray pulsars listed
in Table 2.\\
The results plotted in Figure 3, apart from a normalization factor,
come therefore from
\begin{equation}
C_{norm}(E_1,E_2) = \frac{E^{2.1}}{C_{Crab}(E_1,E_2)} \times
\frac{C_{source}(E_1,E_2)}{Continuum(E)}
\end{equation}
where $C_{Crab}$ and $C_{source}$ are the Crab nebula and the source
count rate spectra respectively and $Continuum$ is the functional form
of the continuum of $source$ from the fit.\\
In this plot the cyclotron features listed in Table 2 are apparent. In
4U 1907+09 and Vela X--1 the possible fundamental at half the line
energy is barely visible. It is also clearly evident the difference
between the sources showing a cyclotron feature in their spectrum (the
first five from top) and the other two (GX1+4 and GS1843+00) that do not
show any feature, included for comparison.

Confining ourselves to a purely phenomenological point of view we
therefore conclude that in absence of a reliable theoretical model 
all claims of presence of a double feature in the spectrum of a source
should be substantiated by a clear detection also in its normalized Crab
ratio. While this second method is less sensitive, it is rather robust,
as it is in first approximation independent from instrument calibrations
and, using only the non-normalized Crab ratio, even from any
modelization of the source continuum.\\
A caution must be used while having a negative result, as this method is
only qualitative. Given that the detectors are relatively dispersive and
that the fundamental line is broad and with a small equivalent width, no
definitive statement can be said in the case of apparent non--detection,
unless some quantitative test is used.

\centerline{\psfig{file=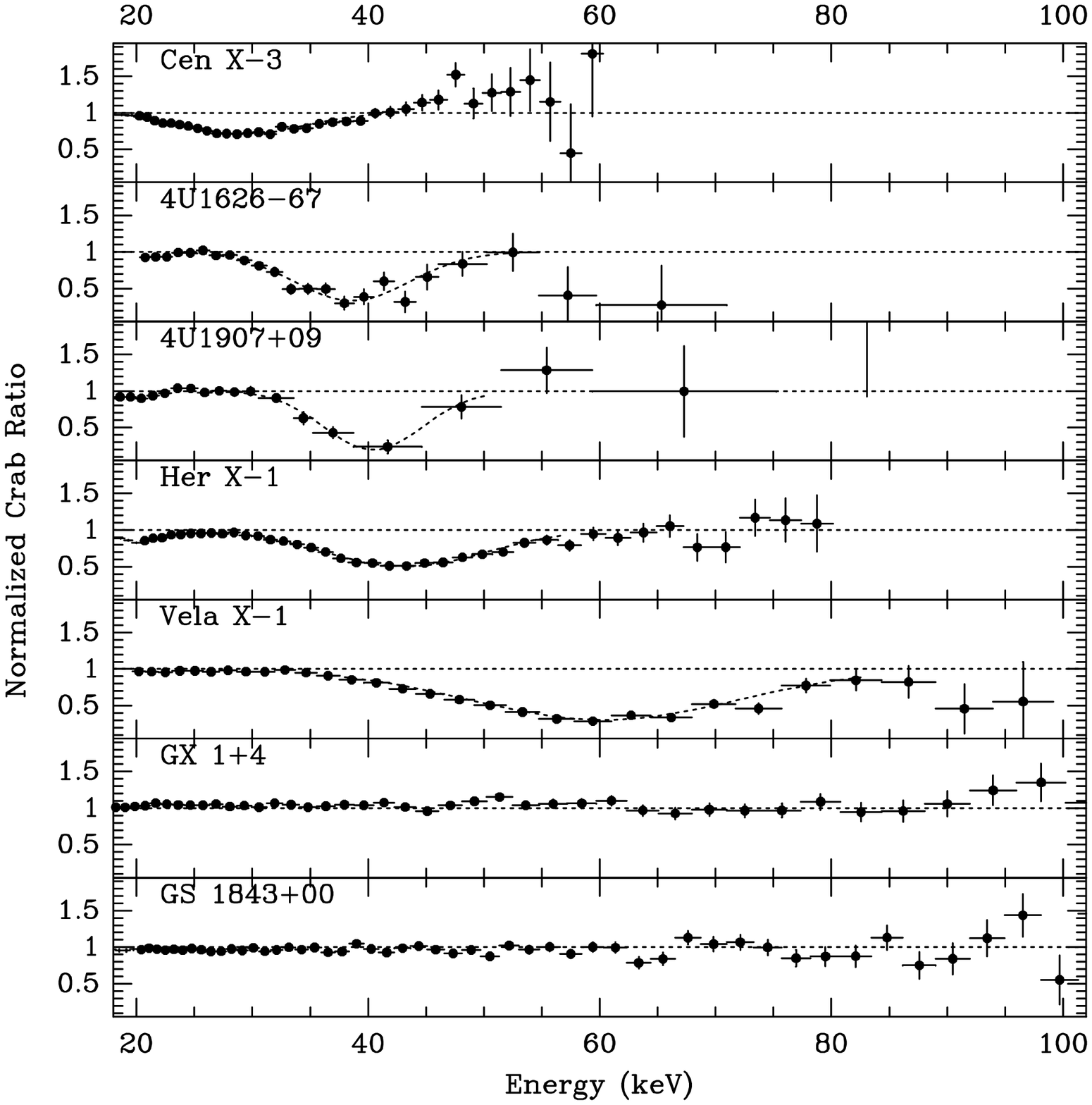,height=10truecm}}
\vspace{-0.7truecm}
\begin{center}
\begin{minipage}[t]{0.8\textwidth}
{\bf Figure 3:\rm the normalized Crab ratio of seven
X--ray pulsars as observed by BeppoSAX. See text for details}
\end{minipage}
\end{center}

\section{DISCUSSION}

The detections of the cyclotron line features give direct estimates of
the intensity of the magnetic field of the neutron stars at the emission
zone, that should be near the neutron star surface, apart the case
of extreme radiative shock.

The measured resonance energies of course must be corrected for the
gravitational redshift due to relativistic gravitational field of the
neutron star. The correction factor $\frac{1}{1+z}$ depends on the
square root of $\frac{M}{R}$, where M is the mass of the neutron star
and R is the radial distance from the centre of the neutron star of the
emission zone. We can reasonably assume that R$\sim R_{NS}$, that is the
emission zone is at the surface of the neutron star or negligibly above
it, but this assumption may break if a radiative shock is present above the
neutron star, dislocating the emission region at a
non negligible distance from the NS surface.\\
The other possible uncertainty
in this correction is the mass of the neutron star itself and its
mass/radius ratio. A Monte Carlo method applied to orbital data
(Rappaport and Joss 1983) shows that NS masses should be between 1 and 2
solar masses. If we assume a 50\% possible scatter in the NS masses and
if we assume a value of R using the more recent equation of states
coupled with the constraints coming from the observation of kHz QPOs in
LMXRBs (Miller, Lamb and Cook
1998), we have an additional $\pm$10\% scatter in the {\it corrected}
resonance energy, to be added to the statistical uncertainty from the
fit, that usually is of the order of 1--3\%. The observed resonance
energy $ E_{cyc}^\infty$ must be corrected using equation 2. 

\begin{equation}
E_{cyc}^0 = E_{cyc}^\infty \times \left( 1 - 0.295 \frac{M}{M_\odot}
\frac{10{km}}{R} \right)^{-\frac{1}{2}}
\end{equation}

A simple interpretation of the observed line broadening is that it is
produced via thermal Doppler broadening (e.g. M\'esz\'aros 1992). In
this case the FWHM of the feature can be a measure of the temperature of
the emitting atmosphere using equation 3.

\begin{equation}
\Delta \omega_B \simeq \omega_B \left( 8 \times \ln(2) \times
\frac{\rm kT_e}{\rm m_ec^2} \right)^{\frac{1}{2}} |\cos\theta|
\end{equation}

It is reasonable that different sources with different intrinsic
luminosities have atmospheres with different temperatures (e.g.
Harding et al. 1984). Moreover equation 3 depends on the
average cosine of the aspect angle $\theta$ (the angle between the
magnetic field axis and the line of sight). However the
data in Table 2 show a correlation between the centroid energy and
the width of the measured cyclotron features, even if with some scatter.
This correlation is plotted in Figure 4, in which we added the data coming
from the OSSE measurement of the cyclotron feature in A0535+26 during
its 1994 outburst (Grove et al. 1995).

\centerline{\psfig{file=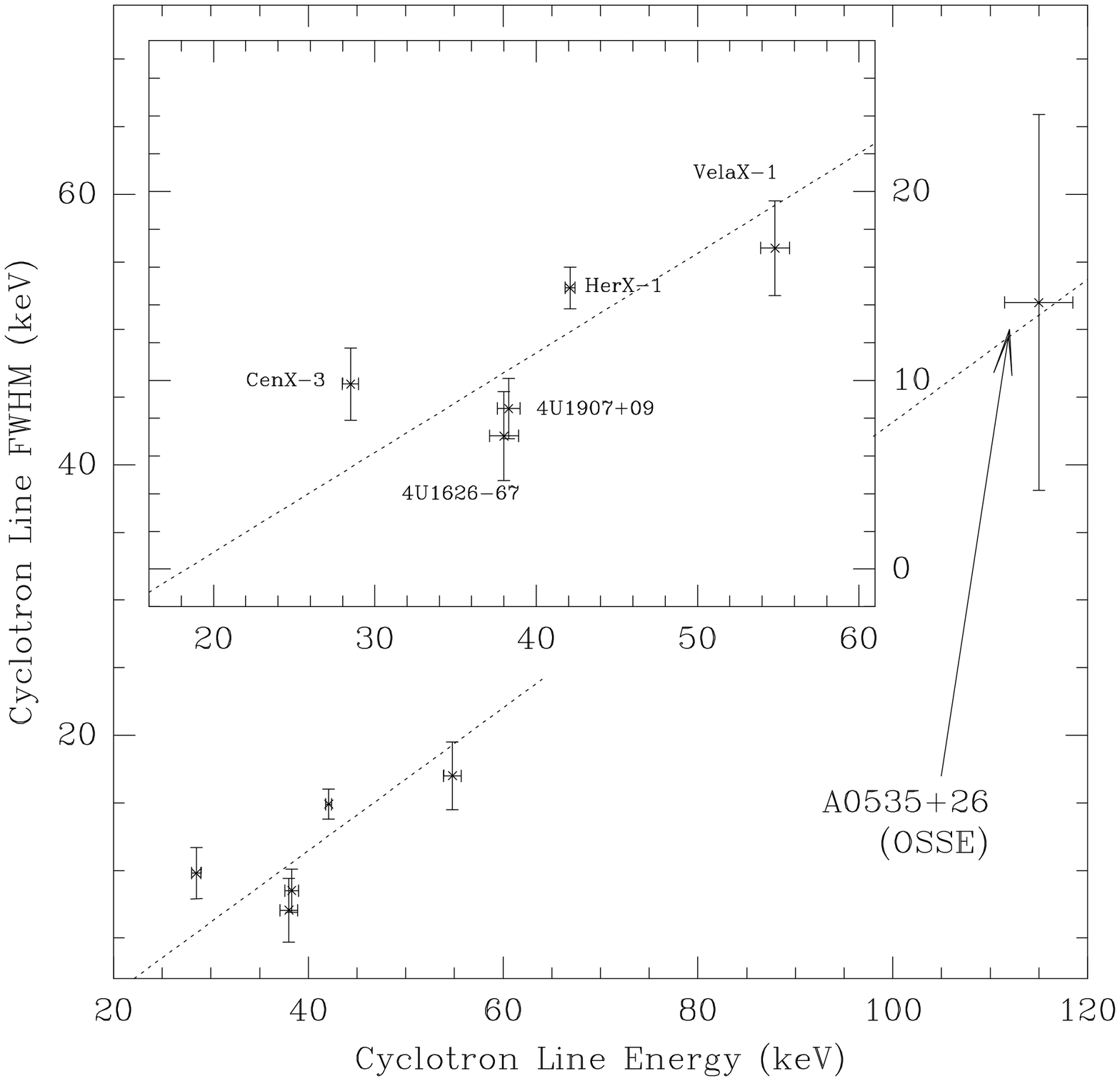,height=10truecm}}
\vspace{-0.7cm}
\begin{center}
\begin{minipage}[t]{0.8\textwidth}
{\bf Figure 4:\rm FWHM versus centroid energy for the cyclotron features
measured with BeppoSAX. The point for A0535+26 comes from the OSSE
measurement during the 1994 outburst (Grove et al. 1995)}
\end{minipage}
\end{center}

Therefore, from this apparent correlation one can conclude that the
spread in both mass/radius ratio (this is {\it not} the NS mass/radius
ratio, as the emission zone may be relatively distant from the NS
surface as discussed above) and in the temperature of the
atmosphere is relatively small. The estimated temperatures of the
atmospheres are reported in Table 3, using equation 3 as a first order
approximation. The estimates of NS masses are also reported, when
available (Rappaport and Joss, 1983).

\begin{center}
{\bf Table 3}

\begin{tabular}{|llllll|}
\hline
Source name & Centroid energy & FWHM & kT$_e$ & NS mass$^1$& B field\\
            & (keV) & (keV) & (keV) & (M$_\odot$) & at surface \\
	    & & & & & (10$^{12}$ G) \\
\hline
 & & & & & \\
 Cen X--3 & 27.9$\pm$ 0.5 & 9.8$\pm$ 1.9& 11.3$\pm$4.5 &
 1.07$^{+0.63}_{-0.57}$ & 3 \\
 4U 1626--67 & 37$\pm$1 & 7.05$\pm$2.35 & 3.3$\pm$2.1&  & 4.1 \\
 4U 1907+09 & 39.4$\pm$0.6 & 8.5$\pm$1.6& 4.5$\pm$1.6&  & 4.4 \\
 Her X--1 & 42.1$\pm$0.3 & 14.9$^{+1.25}_{-1.0}$&11.3$\pm$1.9&
 1.45$^{+0.35}_{-0.40}$  & 4.8 \\
 Vela X--1 & 54.4$^{+1.5}_{-0.2}$&17$^{+3}_{-2}$ & 8.9$\pm$2.9&
 1.85$^{+0.35}_{-0.30}$& 7.0\\
 A0535+26 & 115$^{+3}_{-4}$ & 52$^{+13.8}_{-14}$ & 18.5$\pm$10 & &
 13\\
\hline
\multicolumn{6}{l}{ }\\
\multicolumn{6}{l}{$^{\mbox{1}}$ NS mass is assumed 1.4 M$_\odot$ when an
estimate is not available}\\
\end{tabular}
\end{center}

From this table some spread in the estimated electron temperature is
evident. This may be intrinsic, but we caution that we used a rough
estimate to obtain these values. More reliable values will certainly
come from a self--consistent model of X--ray pulsars atmosphere, when
available for data fitting. These values are in fair agreement with the
self--consistent calculations of Harding et al. (1984).

As previously noted, in at least two cases (4U 1907+09 and Vela X--1)
the observational data may show, depending on the model used, a double
cyclotron feature. The resonant photon energies are
\begin{equation}
\hbar\omega = \frac{\left[ -m_ec^2 + \left( m_e^2c^4 + 2m_ec^2
\ \hbar\omega_c\ n \ \sin^2\theta\right)^\frac{1}{2} \right]}{\sin^2\theta}
\end{equation}

where $\hbar\omega_c$ is the fundamental resonance energy and {\it n} is
the Landau principal quantum number. As can be seen, the levels are not
exactly equispaced.\\
The possible presence of the first harmonic of the fundamental frequency
was discussed e. g. by M\'esz\'aros and Alexander (1989) in the
framework of Gamma Ray Bursts and by Araya and Harding (1995) for
A0535+26. A naive approach, as suggested by M\'esz\'aros and Alexander,
should bring to the conclusion that the line at the lower energy, the
fundamental, should be deeper given that the cyclotron opacity decreases
for increasing harmonics. Actually the observational scenario, even if
with a non negligible ambiguity, suggest that in the case of two
features, the first harmonic is much deeper than the fundamental,
exactly the contrary than that said before.

A way around this problem may be the addition to the radiation transfer
of two--photon scattering (Alexander and M\'esz\'aros, 1991;
M\'esz\'aros, 1992). 
Due to this process, the photons are redistributed from the higher
harmonics to the lower. A photon with energy 2$\omega_c$ therefore may
be splitted in two photons of energy $\omega_c$.
The addition of this process to the calculations
gives much deeper first harmonic, and, parenthetically, also a second
harmonic. Of course, in order that the photon splitting of this process
be sufficiently effective to replenish the fundamental feature,
enough photons at and above 2$\omega_c$ must be present. This may be a
problem with X--ray pulsar spectra.

In the case of the possible double lines in A0535+26 (Grove et al. 1995)
Araya and Harding (1996) calculated X--ray spectra near the fundamental
and the first harmonic for both a ``low'' (5.2$\times 10^{12}$G) and a
``high'' (10.7$\times 10^{12}$G) magnetic field. The difficulty to model
both the low energy (55 keV) and the high energy (115 keV) features with
their model brings the authors to the conclusion that the ``high'' field
hypothesis must be preferred. This obviously implies that the 115 keV
observed feature corresponds to the fundamental resonance energy of the
magnetic field in A0535+26.

A different approach is suggested by Alexander et al. (1995). They
suggest that the double features in the spectra of X--ray pulsars may be
produced in separate parts of a shocked accretion column and that one of the
features is produced via Doppler shifting by the pre--shock plasma.
In this case, the ``blue'' line is the fundamental and the ``red'' one
is due to the interaction of the emitted radiation with the pre--shock
infalling matter, that has relativistic velocity. In this case, the
infalling electron ``see'' the photons emitted by the post-shock high
density atmospheres as blue shifted. Therefore the resonant scattering
does not occur anymore at the energy $\hbar\omega_c$, but at a lower
energy, depending on the velocity of the infalling electrons.

In summary, this report shows that the BeppoSAX campaign on X--ray
pulsars allowed to extend the observational results on cyclotron lines,
giving a small but significant set of measures of fundamental physical
quantities of accreting magnetized neutron stars.

\section{REFERENCES}

\vspace{-5mm}

\setlength{\itemindent}{-8mm} \setlength{\itemsep}{-3mm}

\begin{itemize}
\item Alexander, S. G., M\'esz\'aros, P., Astrophys. J. Letters, {\bf
344}, L1 (1989)
\item Alexander, S. G., M\'esz\'aros, P., Astrophys. J. Letters, {\bf
372}, 554 (1991)
\item Alexander, S. G. \etal, Astrophys. J., {\bf 459}, 666 (1995)
\item Araya, R. A., Harding, A. K., Astrophys. J. Letters, {\bf 463},
L33 (1996)
\item Cusumano, G. \etal, Astron. Astrophys., {\bf 338}, L79 (1998)
\item Dal Fiume, D. \etal, Astron. Astrophys., {\bf 329}, L41 (1998)
\item Davidson, K. and Ostriker, J. P., Astrophys. J., {\bf 179}, 55 (1973)
\item Grove, J. E. \etal, Astrophys. J. Letters, {\bf 438}, L25 (1995)
\item Harding A. K. \etal, Astrophys. J., {\bf 278}, 369 (1984)
\item Israel, G. \etal, Nucl. Phys. B (Proc. Suppl.), {\bf 69}, 141
(1998)
\item Makishima, K. \etal, Astrophys. J. Letters, {\bf 365}, L59 (1990)
\item Mihara, T. \etal, Astrophys. J. Letters, {\bf 379}, L61 (1991)
\item Mihara, T., Ph D. Thesis, Univ. of Tokyo, RIKEN IPCR CR--76 (1995)
\item Miller, M. C. \etal, Astrophys. J., {\bf 509}, 793 (1998)
\item M\'esz\'aros, P., {\it ``High--Energy Radiation from Magnetized
Neutron Stars''}, The University of Chicago Press, Chicago (1992)
\item Nagase, F. \etal, Astrophys. J. Letters, {\bf 375}, L49 (1991)
\item Orlandini, M. \etal, Astrophys. J. Letters, {\bf 500}, L165 (1998a)
\item Orlandini, M. \etal, Astron. Astrophys., {\bf 332}, 121 (1998b)
\item Piraino, S. \etal,  Nucl. Phys. B (Proc. Suppl.), {\bf 69}, 220
(1998)
\item Piraino, S. \etal, Astron. Astrophys. (submitted) (1999)
\item Rappaport, S. A. and Joss, P. C., in {\it ``Accretion--driven
stellar X--ray sources''}, W. H. G. Lewin and E. P. J. van den Heuvel
editors, Cambridge University Press, Cambridge, p. 1 (1983)
\item Santangelo, A. \etal, Astron. Astrophys., {\bf 340}, L55 (1998)
\item White, N. E. \etal, Astrophys. J., {\bf 270}, 711 (1983)
\end{itemize}
\end{document}